\documentclass{rspublic}
\usepackage{graphicx}
\newcommand{\pderiv}[2]{\ensuremath{\frac{\partial#1}{\partial#2}}  } 

\renewcommand{\exp}[1]{e^{#1}}

\newcommand{\vect}[1]{\ensuremath{\mathbf{#1}}}

\newcommand{\EE}{\ensuremath{\mathbf{E}}}
\newcommand{\DD}{\ensuremath{\mathbf{D}}}
\newcommand{\BB}{\ensuremath{\mathbf{B}}}
\newcommand{\HH}{\ensuremath{\mathbf{H}}}

\newcommand{\vel}{\ensuremath{\mathbf{v}}}

\newcommand{\X}{\ensuremath{\hat{\chi}}}
\newcommand{\Xt}{\ensuremath{\hat{\chi}^{T}}}
\newcommand{\sig}{\ensuremath{\hat{\sigma}}}

\newcommand{\chixy}{\ensuremath{\chi_{xy}} }
\newcommand{\chiyx}{\ensuremath{\chi_{yx}} }
\newcommand{\Dchi}{\ensuremath{\Delta\chi} }

\newcommand{\kk}{\ensuremath{\mathbf{k}}}
\newcommand{\ee}{\ensuremath{\mathbf{e}}}

\newcommand{\gd}{\ensuremath{\mathbf{g}}}

\newcommand{\avg}[1]{\ensuremath{\langle #1 \rangle}}

\newcommand{\curl}{\nabla\times}

\newcommand{\Lag}{\ensuremath{\mathcal{L}}}

\newcommand{\etal}{{\it et al.}~}

\newcommand{\fig}[1]{Fig. \ref{#1}}  
\newcommand{\tab}[1]{Table \ref{#1}} 

\newcommand{\sect}[1]{Section \ref{#1}}   

\newcommand{\ut}[1]{\,\rm{#1}}  

\usepackage{cancel}

\begin{document}


\title[Simpler derivation of Feigel effect and experiments]{Alternative derivation of the Feigel effect and call for its experimental verification}

\author[O. A. Croze]{Ottavio A. Croze}
\affiliation{School of Mathematics and Statistics, University of Glasgow, Glasgow G12 8QW, U.K.
}
\label{firstpage}
\maketitle

\begin{abstract}{Feigel effect, quantum vacuum, magnetoelectric, strong magnetic fields, particle tracking velocimetry, vacuum radiometer}
A recent theory by Feigel [Phys. Rev. Lett. {\bf 92}, 020404 (2004)] predicts the finite transfer of momentum from the quantum vacuum to a fluid placed in strong perpendicular electric and magnetic fields. The momentum transfer arises because of the optically anisotropic magnetoelectric response induced in the fluid by the fields. After summarising Feigel's original assumptions and derivation (corrected of trivial mistakes), we rederive the same result by a simpler route, validating Feigel's semi-classical approach. We then derive the stress exerted by the vacuum on the fluid which, if the Feigel hypothesis is correct, should induce a Poiseuille flow in a tube with maximum speed $\approx 100\mu$m/s ($2000$ times larger than Feigel's original prediction). An experiment is suggested to test this prediction for an organometallic fluid in a tube passing through the bore of a high strength magnet. The predicted flow can be measured directly by tracking microscopy or indirectly by measuring the flow rate ($\approx 1$ml/min) corresponding to the Poiseuille flow. A second experiment is also proposed whereby a `vacuum radiometer' is used to test a recent prediction that the net force on a magnetoelectric slab in the vacuum should be zero.
\end{abstract}


\section{Introduction}

It is well known that quantum vacuum fluctuations can transfer momentum to macroscopic matter. The momentum transfer usually results from a modification the spectrum of allowed vacuum modes by symmetric boundaries. For example, in the Casimir effect (Casimir 1948; Lamoreaux 1997, 2005; Milonni 1994) two conducting plates are placed parallel to each other. The plated reduce of the allowed vacuum modes between them, which causes the total pressure (due to the momentum transfer of vacuum modes) to be smaller between the plates than outside, resulting in a net attraction between the plates: the  Casimir force (Milonni {\it et al} 1988).

However, the vacuum does not usually transfer its momentum to matter in the absence of boundaries: because its fluctuation spectrum is isotropic in free space the expectation value of the momentum density of the vacuum is zero (Milonni 1994). In this paper we revisit a theoretical argument proposed recently (Feigel 2004), which claims that it is possible to transfer momentum to an isolated region of dielectric liquid if it is placed in perpendicular crossed electric and magnetic fields. Critical to Feigel's argument is the assumption that vacuum fluctuations see the dielectric in crossed fields as an anisotropic magnetoelectric medium. This implies vacuum modes will propagate faster (have greater momentum) in one direction than in the opposite direction, causing a net momentum transfer to the dielectric liquid.

In going through Feigel's original argument we have corrected trivial errors in the original derivation. The corrected expressions agree with our own more direct derivation of Feigel's result. In the original paper Feigel proposes a prediction for the speed of a fluid as a result of the momentum transfer from the vacuum. The prediction was, however, predicated on an ideal fluid, and thus not realistically testable. We derive here the flow of a real, viscous fluid driven into motion by the stress on the fluid caused by the vacuum. An experimental test of this improved prediction is proposed. Indeed, this paper is written in the hope that our revised predictions will stimulate experiments to try and measure Feigel's original effect and related recently proposed predictions. Since the publication of Feigel's result, several theoretical works have emerged which have been very enthusiastic about Feigel's idea (van Tiggelen {\it et al} 2005,  van Tiggelen {\it et al} 2006, Shen {\it et al} 2006, Birkeland \& Brevik 2007, Obukhov \& Hehl 2008). Some have, however, questioned the soundness of the original argument stating that no unbounded macroscopic Feigel effect should exist if proper regularisation is applied to the momentum integral (van Tiggelen {\it et al} 2005,  van Tiggelen {\it et al} 2006). Microscopically, however, recent work suggests that a properly regularised Feigel effect could exist (Kawka  {\it et al} 2010). Most other alternative macroscopic theories consider the momentum from the vacuum for magnetoelectric fluids (Birkeland \& Brevik 2007) and samples of magnetoelectric materials (van Tiggelen {\it et al} 2006) confined in parallel plate geometries, like in the Casimir effect. The latter work predicts an unmeasurably small linear momentum transfer (van Tiggelen {\it et al} 2006). Interestingly, following a semi-classical approach similar to Feigel, Obukhov \& Hehl (2008) found that there is no net force on a magnetoelectric slab of finite thickness in the vacuum. Feigel (2009) has also recently considered the interesting possibility of constructing `quantum wheels' using magnetoelectric nanoparticles. No experiments have been carried out to test whether Obukhov \& Hehl's null result, which implies Feigel's wheels should not function. More surprisingly, Feigel's original theory, which remains the one predicting the largest effect, remains untested. Experimental tests for the effects described are suggested at the end of the paper, but we leave the field open to imaginative experimentalists with access to the strong fields or magnetoelectric materials required. 

The paper is organised as follows: in \sect{Model} the physical set--up for the model is presented together with quick summary of the derivation and underlying assumptions. In \sect{simplerroute} our quicker route to Feigel's quantum result is then presented; in \sect{Predictions} a new expression for vacuum stress is derived and a revised prediction of Feigel's theory is evaluated and discussed together with Obukhov \& Hehl's null prediction; in \sect{Experiments} we make some suggestions for experimentally testing Feigel's  and Obukhov \& Hehl's predictions. Finally, in \sect{Discussion} conclusions are drawn from the preceding analysis and a few final comments are made.


\section{Feigel's semi-classical model \label{Model}}

We summarise here Feigel's semi--classical derivation and assumptions (highlighted in italics). Slight inaccuracies in the derivation of two key results have been corrected, so these expressions differ from those in Feigel's original paper (a table comparing the original results with our corrected ones is presented in \ref{AppdxC}).

\subsection{Physical system and initial conditions}

Feigel considers the following situation: a region of a dielectric fluid far
from the boundaries of its container is initially at rest ($t=0$). Subsequently strong
electric and magnetic fields crossed at right angles to each other are applied
to the region. As the fields reach their constant final values,
$\EE_{\ut{ext}}$ and $\BB_{\ut{ext}}$ for electric and magnetic fields
respectively, the fluid is accelerated by Lorentz forces ($F_{\ut{Lorentz}}\propto\partial_t(
\EE_{\ut{ext}}\times\BB_{\ut{ext}})$) to a final velocity $\vel$.
%

\subsection{Brief summary of Feigel's derivation and fundamental assumptions \label{Assumptions}}

\begin{enumerate}
 \item {\it Feigel assumes that the portion of dielectric fluid under consideration is to a good approximation 
 ideal (inviscid), incompressible, homogeneous and not
acted upon by external stresses or body forces}; 
\item as the fields become steady, they have a maximum momentum, the opposite of which is transferred to the fluid 
by momentum conservation, since the combined system conserves its initial zero net momentum. Feigel derives 
the conservation law from the relativistically transformed Lagrangian of the moving 
dielectric. This provides the `classical' fluid momentum (ignoring terms of order $(v/c)^2$)
\begin{equation}\label{classical}
\rho \vel = \frac{\epsilon\mu -1 }{4\pi \mu c}(\EE\times\BB)
\end{equation}
It should be noted that {\it this conservation law holds only if the region under consideration is stress free}, as mentioned in point $1$. Otherwise momentum is not conserved.
\item It has been shown that the optical response of a dielectric in crossed $\EE$ and $\BB$ fields is
the same as that of a magnetoelectric material (Roth \& Rikken 2002). {\it Feigel assumes that electromagnetic modes of the vacuum will also `see' a 
magnetoelectric}. The Lagrangian density of a magnetoelectric material is derived by Feigel by relativistic transformation of the Lagrangian of a magnetoelectric (see equation (\ref{EMLagMedia4}) of \ref{AppdxB}) in the small speed limit (ignoring, as above, terms of order $(v/c)^2$). The Euler-Lagrange equation of this Lagrangian then provides a momentum conservation law:
\begin{equation}\label{MElagrangian}
\rho \vel=\frac{1}{4\pi} \left( \frac{\epsilon \mu -1}{\mu c}
\EE\times\BB+\frac{1}{\mu c} \EE\times(\Xt \EE) + \frac{1}{\mu c}
(\X\BB)\times\BB\right)
\end{equation}
which allows to evaluate the momentum associated with the vacuum. We note, that using the inverted constitutive equations (\ref{Con11}) and (\ref{Con22}), we can rewrite the above as: 
\begin{equation}\label{MElagrangian22}
\rho \vel=\frac{1}{4\pi c} \left(\DD\times\BB-\EE\times\HH\right)
\end{equation}
we will consider the appropriateness of using this momentum density to evaluate the fluid momentum in the Discussion.
\item The field modes and refractive indices for electromagnetic waves in a magnetoelectric are given by (taking the optical axis along the $\ee_{3}$ direction, see \ref{AppdxA}),
\begin{equation*}\label{MEmodes000}
[\EE_{\pm\kk1}, \BB_{\pm\kk1}]=E_{\pm\kk1}[\ee_{1}, n_{\pm\kk1}\ee_{2}],\,\,\,\,\,\, [\EE_{\pm\kk2}, \BB_{\pm\kk2}]=E_{\pm\kk2}[\ee_{2}, -n_{\pm\kk2}\ee_{1}]
\end{equation*}
where $E_{\pm\kk\lambda}=E_{0 k}\exp{(k_\lambda z-\omega t)}$, for each polarisation $\lambda=1,2$, and
\begin{equation*}
n_{\pm\kk,1}=\pm n_0+\chixy,\,\,\,\,\,\,\,n_{\pm\kk,2}=\pm n_0-\chiyx
\end{equation*}
where $\chixy$ and $\chiyx$ are the magnetoelectric susceptibilities responsible for the optical anisotropy of the magnetoelectric (see \ref{AppdxA}). Feigel substitutes the above modes and refractive indices into (\ref{MElagrangian}) to evaluate the time-averaged (denoted by the overbar) momentum flux for a mode in the $z$-direction:
\begin{equation}\label{MomDens}
\overline{\rho v}_k=2\Delta\chi \frac{1}{c}\frac{\epsilon E^2_{0 k}}{4\pi}
\end{equation}
where $\Delta\chi\equiv\chixy-\chiyx$.
\item Feigel then replaces the electric field with its operators and {\it evaluates the electromagnetic vacuum energy density expectation value}
\begin{equation*}
\avg{0|\frac{\epsilon \hat{E}^2_{0 k}}{4\pi}|0}=\frac{1}{V}\frac{1}{2}\hbar \omega
\end{equation*}
to obtain the vacuum momentum density per mode from (\ref{MomDens}):
\begin{equation}\label{VacMomentumDens0}
g_{0 k}\equiv\avg{0|\overline{\rho v}_k|0}= \frac{1}{V}\Dchi \hbar k_0,
\end{equation}
where $k0\equiv \omega/c$.
\item Summing over all modes the total momentum density in the z-direction is then:
\begin{equation}\label{VacMomentumDens001}
g_0\equiv\sum_{k} g_{0 k}=\frac{1}{V}\Dchi \hbar \sum_{k_0}k_0\to\frac{1}{2 \pi^2} \Delta \chi \hbar \int_0^\infty k_0^3 d k_0.
\end{equation}
where the last step involves the standard replacement $\sum_{k_0}\rightarrow\frac{V}{8 \pi^3}\int d^3 k_0$ (see next section). 
\item The integral in (\ref{VacMomentumDens001}) is divergent. Feigel makes the {\it crucial assumption that vacuum modes with frequency greater than the dielectric's ``cutoff frequency'', $\omega_{\ut{c}}$ (the frequency above which the
dielectric's molecular polarisability vanishes) do not interact with it}. (Implicit in the derivation is also the assumption that absorption and dispersion are not significant, i.e. that for $\omega<\omega_{\ut{c}}$, the permittivity and magnetoelectric susceptibility of the dielectric as seen by vacuum modes does not change appreciably with frequency.). This allows to
evaluate a finite value for the momentum density (\ref{VacMomentumDens001}):
\begin{equation}\label{VacMomentum5}
g_0=\rho\,v_{\ut{vac}}=\frac{1}{8 \pi^2} \Delta \chi  \frac{\hbar \omega_{c}^4}{c^4}.
\end{equation}
From which, dividing by the fluid density $\rho$, Feigel obtained an estimate for the vacuum contribution to the fluid speed, $v_{\ut{vac}}$. We iterate that the above expression is different from that obtained by Feigel due to some trivial errors in the original derivation (see  \ref{AppdxB}).
\item The magnitude of the corresponding ``classical'' contribution of the dielectric fluid's speed is given by (\ref{classical})
\begin{equation}\label{MomentumCons}
v_{\ut{class}}=
\frac{1}{\rho}\frac{\epsilon\mu -1 }{4\pi \mu c}E_{\ut{ext}} B_{\ut{ext}}.
\end{equation}
the relative magnitude of quantum vacuum and classical contributions will be discussed in section \ref{Predictions}.
\end{enumerate}

\section{Simpler derivation of the vacuum contribution to magnetoelectric momentum density\label{simplerroute}}

Here we present our alternative route to Feigel's result, equivalent to steps $1-5$, but very much quicker. From quantum electrodynamics the expectation value for the momentum density $\hat{\gd}$ of the vacuum is given by (Milonni 1994):
\begin{equation}\label{VacMomentumDens01}
\gd_0\equiv\avg{0|\hat{\gd}|0}= \frac{1}{V}\sum_{\kk\lambda}\frac{1}{2}\hbar\kk,
\end{equation}
where $V$ is a sample volume of the medium under consideration and $\kk$ is the wave vector of each vacuum mode {\it in this medium} with possible polarisation states $\lambda=1,2$. Next we assume with Feigel that the vacuum experiences the same birefringence in a magnetoelectric medium as light does. The medium parallel to the optical axis  in the $\ee_z$ direction has the following dispersion relation:
\begin{equation}\label{VacDispRel}
\kk\cdot\ee_z= k_0 n_{\kk\lambda},
\end{equation}
where $n_{\kk\lambda}$ (given below) are the refractive indices parallel to the optical axis and we define $k_0=\omega/c$. On the other hand, in directions perpendicular to the optical axis the medium is isotropic, so that contributions to (\ref{VacMomentumDens01}) vanish by symmetry. Substituting in (\ref{VacDispRel}) then reduces to:
\begin{equation}\label{VacMomentumDens02}
\gd_0= \frac{1}{V} \frac{1}{2}\hbar \sum_{k_0}\sum_\lambda k_0 \left(n_{+\kk\lambda}+n_{-\kk\lambda}\right) \ee_z,
\end{equation}
where the sum has been expanded in terms of the contributions by modes for each direction ($\pm$) of travel along $\ee_z$. The expressions for the anisotropic indices $n_{\pm\kk\lambda}$, derived in \ref{AppdxA}, are:
\begin{equation}\label{MEri000}
n_{\pm\kk,1}=\pm n_0+\chixy,\,\,\,\,\,\,\,n_{\pm\kk,2}=\pm n_0-\chiyx,
\end{equation}
where we recall $|\chi_{ij}| \ll|n_0|$ so, e.g., $n_{\kk,1}>0$ and $n_{-\kk,1}<0$. Substituting these expressions
into (\ref{VacMomentumDens02}) and summing over all polarisations $\lambda$ gives (considering only the magnitude of $\gd_0$):
\begin{equation}\label{VacMomentumDens03}
g_0= \frac{1}{V} \Delta \chi \hbar \sum_{k_0} k_0,
\end{equation}
where $\Delta \chi\equiv \chixy-\chiyx$. Next, making the standard replacement $\sum_{k_0}\rightarrow\frac{V}{8 \pi^3}\int d^3 k_0$, we find:
\begin{equation}\label{VacMomentumDens04}
g_0= \frac{1}{2 \pi^2} \Delta \chi \hbar \int_0^\infty k_0^3 d k_0.
\end{equation}
These integration limits assume modes of all wavelengths contribute to the momentum density, as they would in free space, so $g_0$ is divergent. However,  vacuum electromagnetic modes with very small wavelengths are not expected to interact with the macroscopic electromagnetic properties of a material medium. The choice of a reasonable value for the cut-off will be discussed in the next section. Here, with Feigel, we simply assume it is reasonable to approximate (\ref{VacMomentumDens04}) using an upper cut-off on the wavenumber $k_{c}=\omega_c/c$ to finally obtain:
\begin{equation}\label{VacMomentumDens05}
g_0= \frac{1}{8 \pi^2} \Delta \chi  \frac{\hbar \omega_{c}^4}{c^4}.
\end{equation}
Equation (\ref{VacMomentumDens05}) represents the vacuum contribution to the momentum density of a magnetoelectric. In media without the special symmetry of magnetoelectrics the refractive indices obey $n_{+\kk\lambda}=-n_{-\kk\lambda}$, so that $\Delta \chi=0$ and there is no transfer of momentum, as expected. Our alternative derivation provides a concise route to Feigel's result. It is equivalent to his semi-classical approach (once trivial errors in the original derivation are corrected, see \ref{AppdxC}). 

\section{Vacuum stress and realistic predictions \label{Predictions}}

The original prediction of the Feigel effect was obtained assuming conservation of momentum to obtain the vacuum contribution to the speed of a dielectric fluid placed in crossed fields (an effective magnetoelectric) from (\ref{VacMomentumDens05}) (Feigel 2004). Feigel's fluid was ideal and far from boundaries, making the measurement of the vacuum speed as originally predicted a utopian pursuit. We show here how more realistic predictions for the original experiments can simply be obtained and propose a new experimental test. These improved predictions are based on the fact that the transfer of momentum from the vacuum to a magnetoelectric results in a stress. 

For the case of an effective magnetoelectric fluid, an expression for the vacuum stress can be derived applying kinetic theory to a gas of vacuum modes (virtual photons) of momentum $\frac{1}{2}\hbar \kk$ travelling in the fluid. Optical anisotropy implies that the net momentum
\begin{equation}\label{VacMomTran0}
\Delta {\bf p}_\kk= \frac{1}{2}\hbar k_0  \sum_\lambda\left(n_{+\kk\lambda}+n_{-\kk\lambda}\right) \ee_3
\end{equation}
is transferred by counterpropagating vacuum modes across a surface in the fluid of area $A$ and normal to $\ee_z$. The corresponding stress on the fluid is the time rate of change of this momentum transfer per unit area: ${\bf \Pi}_\kk=\frac{1}{A}\frac{\Delta {\bf p}_\kk}{\Delta t}$. Modes crossing $A$ in the interval $\Delta t$ are recruited from a slice of fluid of thickness $\Delta z= c[\frac{1}{n_{+\kk\lambda}}-\frac{1}{n_{-\kk\lambda}}]  \Delta t $ (recall $n_{-\kk\lambda}<0$), where $c$ is the speed of light {\it in vacuo}. Thus, the magnitude of the net stress in the z-direction due to a mode pair is given by:
\begin{equation}\label{VacStress}
\Pi_\kk=\frac{1}{V}\frac{1}{2}\hbar k_0 c \sum_\lambda\frac{n_{-\kk\lambda}^2-n_{+\kk\lambda}^2}{n_{+\kk\lambda}n_{-\kk\lambda}}
\end{equation}
where $V=A \Delta z$. Substituting the refractive indices (\ref{MEri000}) into equation (\ref{VacStress}) (neglecting terms of order $||\X||^2$) and summing over mode magnitudes gives:
\begin{equation}\label{VacStress0}
\Pi_0=  2 \frac{c}{n_0}  \frac{1}{V}\Delta \chi \hbar \sum_{k_0} k_0.
\end{equation}
Comparing this expression with (\ref{VacMomentumDens03}) we see that $\Pi_0= 2 g_0 c/n_0$, as one would expect (Loudon {\it et al} 2005). We can then use (\ref{VacMomentumDens05}) to write: 
\begin{equation}\label{VacStress01}
\Pi_0=4 \pi^2 \Delta \chi \hbar \frac{c}{n_0} \frac{1}{\lambda_c^4}.
\end{equation}
where we recall $n_0=\sqrt{\epsilon \mu}$ is the fluid's isotropic refractive index and note that (\ref{VacStress01}) has been re-written in terms of the cut-off wavelength $\lambda_c$ in view of the evaluations here below. Equation (\ref{VacStress01}) is the magnitude of the stress (acting in the z-direction) exerted by the vacuum on an effectively magnetoelectric fluid. In section (a) below we apply this expression to calculate the speed of a dielectric fluid in a tube with a portion of its length placed in strong crossed fields (a realistic version of Feigel's original scenario). Note that an equivalent result for the vacuum stress could also have been obtained semi-classically by evaluating the contribution from counterpropagating modes to the electromagnetic stress tensor: $\Pi_\kk^{zz}\sim \epsilon E_0^2 (n_{-\kk\lambda}^2-n_{\kk\lambda}^2)$. 

Indeed, it was by using such a semi-classical approach applied to a magnetoelectric slab of finite thickness in a vacuum that Obukhov \& Hehl (2008) recently predicted that the net stress on the slab should be
\begin{equation}\label{VacStress02}
\Pi_0=0.
\end{equation}
That is, the vacuum exerts no net force on a magnetoelectric the slab in spite of its anisotropy! In (b) below we propose to test this prediction with an experiment where the angular drift speed of a `vacuum radiometer' with paddles made of magnetoelectric materials is measured.

\subsection{Dielectric fluid in a tube in crossed fields}

We consider Feigel's original situation of a dielectric fluid placed in perpendicular electric and magnetic fields. In our case, however, the fluid is realistically contained in a tube, a section of which is exposed to the field (see figure \ref{expsetup}a). When the fields are switched on, Lorentz and ponderomotive forces act on the fluid. The fields induce a magnetoelectric susceptibility in the fluid where the fields act, so, according to Feigel's theory, the vacuum exerts a stress in that region. The fluid in the tube thus obeys the following Navier-Stokes equation:
\begin{equation}\label{NS}
\rho_M \frac{\partial{\bf v} }{\partial t}  =  ({\bf P} \cdot \nabla) {\bf E}+\frac{\partial{\bf P} }{\partial t}\times {\bf B}+\eta \nabla^2{\bf v}+ \nabla \cdot {\bf \Pi_0}, 
\end{equation}
where ${\bf v}$ is the flow speed, ${\bf E}$ and ${\bf B}$ are the imposed fields, ${\bf P}=(\epsilon-1) {\bf E}+\X {\bf B}/\mu+o(||\X||^2)$ is the electrical polarisation of the fluid, $\rho_M$ and $\eta$ are the density and kinematic viscosity of the fluid, respectively, and ${\bf \Pi_0}$ is the stress on the fluid due to the vacuum. The fluid is assumed incompressible so $\nabla\cdot{\bf v}=0$ and we note that the advection term $({\bf v}\cdot\nabla){{\bf v}}$ vanishes in the cylindrical geometry of a tube. When both fields and the flow are in the steady-state, and ignoring ponderomotive forces due to edge effects, equation (\ref{NS}) reduces to $\eta \nabla^2{\bf v}+ \nabla \cdot {\bf \Pi_0}=0$. We thus expect a standard Poiseuille flow solution, with a maximum speed, $U_{max}\equiv \max{|{\bf v}|}$, at the center of the pipe given by (Brody \etal 1996): 
\begin{equation}\label{PredicPipe}
U_{max}=\frac{\Pi_0 a^2}{4 \eta L}.
\end{equation}
where $\Pi_0$ is the magnitude of the vacuum stress, $a$ and $L$ are the tube diameter and length, respectively, and $\eta$ is the dynamic viscosity of the fluid, as above. 

To estimate $U_{max}$ we consider, like Feigel, the same organometallic liquids and the same magnitudes of crossed
$\EE$ and $\BB$ fields used in the experiments by Roth and Rikken (Roth \& Rikken 2002). In particular we focus on methylcyclopentadienyl manganese tricarbonyl (MMT), whose relevant properties are shown in \tab{DielLiqProp}.
\begin{table}
\begin{center}
\begin{tabular}{l l l}
  \hline
  Planck's angular constant & $\hbar$ & $1.05\times10^{-34}\ut{J\,s}$ \\
  speed of light & $c$ & $2.99\times10^8\ut{m\,s^{-1}}$ \\
  permittivity of free space & $\epsilon_0$ & $8.85\times10^{-12}\ut{F\,m^{-1}}$ \\
  external electric field & $E_{\ut{ext}}$ & $10^5\ut{V\,m^{-1}}$ \\
  external magnetic field & $B_{\ut{ext}}$ & $17\ut{T}$ \\
  ME susceptibility diff. ($632.8$ nm, room T) & $\Dchi$ & $10^{-11}$ (Roth \& Rikken 2002)\\
  kinematic viscosity ($25^\circ$C) & $\eta$ & $4.5 \times10^{-3}\ut{Pa\,s}$ (YiXing-KaiRun 2011)\\ 
  density ($25^\circ$C) & $\rho_{M}$ & $1.38 \ut{g\,cm^{-3}}$ (Sigma)\\
  refractive index ($589.3$ nm, $20^\circ$C) & $n_0 $ & $1.58$  (Sigma)\\
  cut-off wavelength ({\it cis}-Pt-DEBP) & $\lambda_c$ & $4$ nm (Matassa {\it et al} 2010)\\
  \hline
\end{tabular}
\caption{Values of the parameters used in the evaluation of the vacuum and classical contributions
to the velocity of an effectively-magnetoelectric organometallic fluid.} \label{DielLiqProp}
\end{center}
\end{table}
The magnitude of the vacuum stress is given by (\ref{VacStress01}) and is seen to depend sensitively on the cut-off wavelength $\lambda_c$. Feigel used $\lambda_c\approx 0.1$nm, taking intermolecular distances as a cut-off. However, at such a scales vacuum modes will not interact with a sufficiently large number of molecules to experience magnetoelectric anisotropy. For larger distances, on other hand, modes can interact with the collective effect of an assembly of molecules. The radial distribution function (RDF) for a fluid provides a good measure of the distance beyond which modes see a smooth electromagnetic landscape. The RDF for MMT has not been measured to the best of our knowledge but a recent X-ray diffraction study has measured it for the powder similar organometallic compound, {\it cis}-Pt-DEBP (Matassa {\it et al} 2010). In the study it is found that the RDF becomes flat (uniform density) for $\gtrsim 4$nm, which we assume as the value of the  cut-off $\lambda_c$. The difference in susceptibilities $\Dchi$ is estimated approximately from the value of the birefringence measured by Roth and Rikken experiment (Roth \& Rikken 2002). Using Equations (\ref{MEri000}), we see that this birefringence is $\Delta n\equiv \chixy+\chiyx$ which Roth and Rikken measured as $\Delta n\approx10^{-11}$ for the applied fields shown in \tab{DielLiqProp}. With Feigel we approximate $\Dchi\approx\Delta n$. Hence, using the parameters of \tab{DielLiqProp} equation (\ref{VacStress01}) gives a vacuum stress of $\Pi_0=0.03$ Pa, which using (\ref{PredicPipe}), for a tube with $a=1$ mm and $L=2$ m in (\ref{PredicPipe}), implies a flow with maximum speed 
\begin{equation}\label{predspeed}
U_{max}=100\mu\rm{m/s}.
\end{equation}
This prediction for the flow speed is $4000$ times larger than Feigel's original prediction. The corresponding ``classical'' contribution to the velocity is negligibly smaller, $7.5$ nm/s, as can be seen from (\ref{MomentumCons}) using the parameters in \tab{DielLiqProp}. As a matter of interest we note that (\ref{MomentumCons}) follows from (\ref{NS}) when ponderomotive, vacuum and viscous stress contributions are neglected, so that the classical momentum is conserved. Inclusion of these realistic contributions, however, shows that the classical contribution, which we recall is due to Lorentz forces caused by polarisation currents, is a transient, so it is not only negligible in magnitude, but also at long times when fields are steady. It will not contribute to the steady flow predicted above. 

\subsection{A vacuum radiometer}

Another way to test the reality of vacuum momentum might be to use naturally magnetoelectric compounds. In the organometallic fluids just discussed it is the high fields which induce magnetoelectric anisotropy in the fluid microstructure. However, this anisotropy, and the resulting birefringence, can also arise in solids whose structures support both spontaneous polarisation and magnetisation, breaking both time and space inversion symmetries (Figotin \& Vitebsky 2001). By virtue of this birefringence a slab of such materials in the vacuum should acquire momentum from the vacuum. Now consider an arrangement analogous to a Crookes' radiometer (Crookes 1876, Woodruff 1968), but with the vacuum pressure driving rotation as opposed to temperature gradients: a mill consisting of two or four square panes made of thin magnetoelectric slabs joined together with rods hinged on a low friction axle (see \fig{expsetup}b). If each pane has area $A$ and distance $l$ from the centre of rotation, a vacuum stress of magnitude $\Pi_0$ normal to each pane causes a torque $\tau_0=\Pi_0 A l$. The equation of motion of the radiometer mill is then:
\begin{equation}\label{Radio}
I \dot{\omega}=- \gamma \omega + \tau_0.
\end{equation}
where $\gamma$ is the frictional damping constant and $I=\rho_s A \delta l^2$ its the moment of inertia, where $\rho_s$ is the slab density and $\delta$ its thickness. Integrating equation (\ref{Radio}), we find:
\begin{equation}\label{Radio_sol}
\omega(t)= \omega_\infty(1-e^{-t/t_c}).
\end{equation}
where $\omega_\infty\equiv\tau_0/\gamma$ is the terminal angular speed and $t_c=I/\gamma$ is the characteristic time for the approach to this speed. Assuming the prediction of Obukhov \& Hehl (2008) also applies to slabs of finite extent (i.e. the contributions of any stresses at the edge of the slab are negligible), we then expect
\begin{equation}\label{Radio_sol_OH}
\omega(t)= 0.
\end{equation}
The radiometer should not turn. If in an experiment it did actually turn, measuring its angular drift (see below) would allow to estimate the vacuum stress from 
\begin{equation}\label{Radio_sol_nonzero}
\Pi_0=\frac{\gamma \omega_\infty}{A l}. 
\end{equation}
We would expect such a stress should to scale with $\Delta \chi$, as in (\ref{VacStress01}).

\section{Possible experimental tests \label{Experiments}}

\begin{figure}[tbph]
\centering
\includegraphics[width=0.41\linewidth]{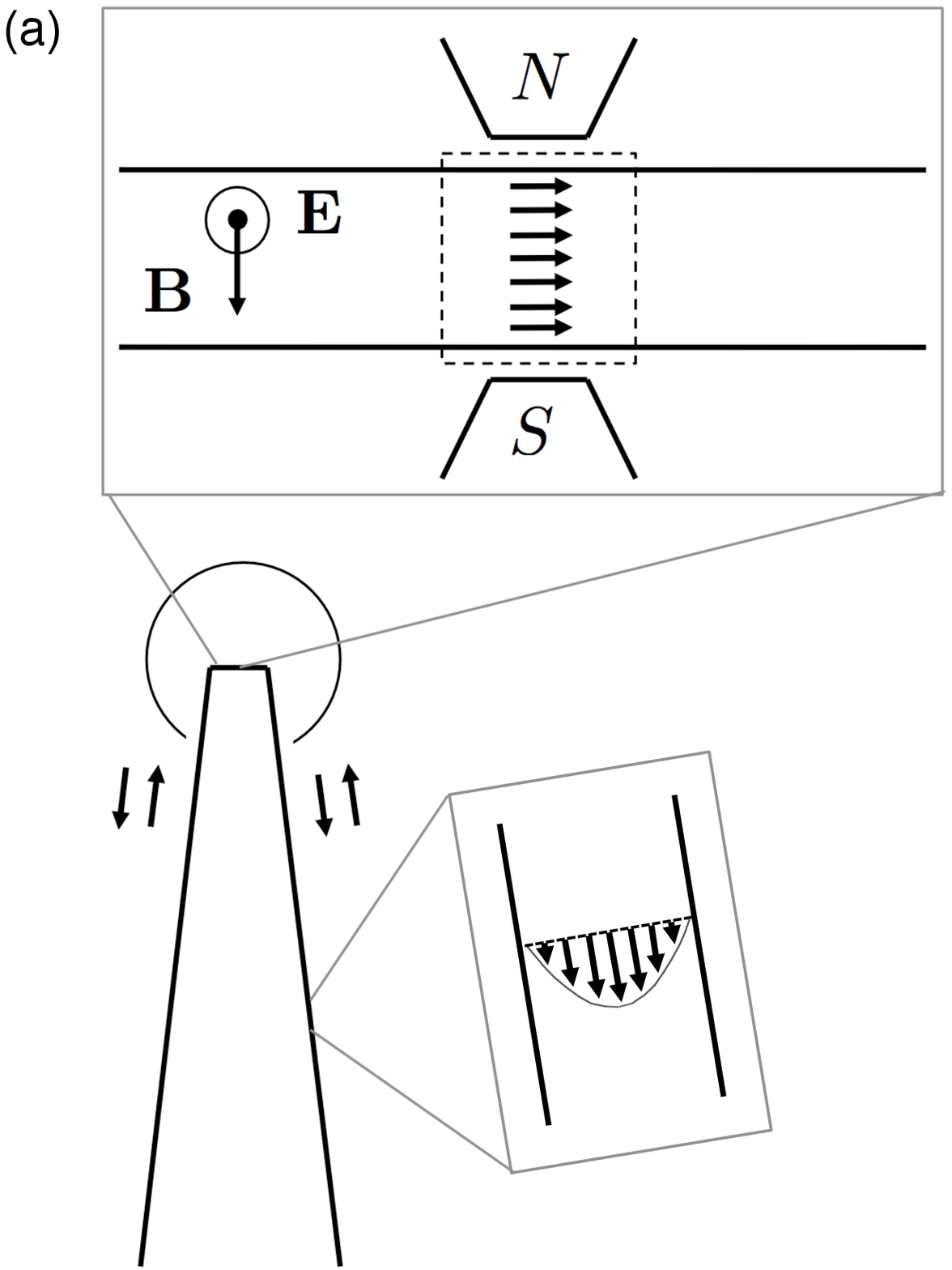}
\includegraphics[width=0.55\linewidth]{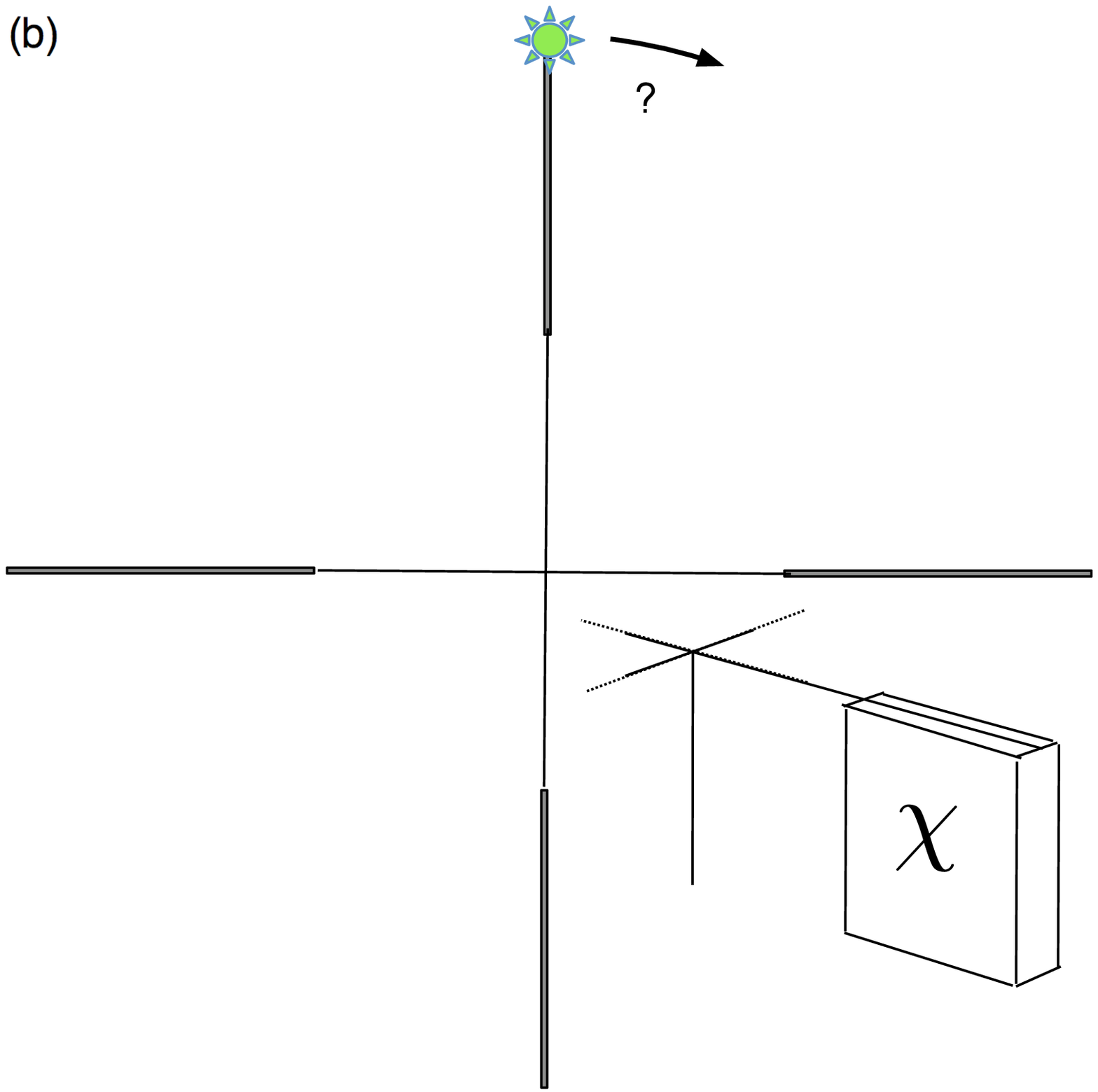}
\caption{Schematic of the proposed experimental arrangements for testing the Feigel effect. (a) A tube is placed in the magnet bore where it is subjected to a strong magnetic field and a perpendicular electric field. The channel enters and exits the magnet via a conical aperture, and flow is measured down stream by the elution or tracking the drift of colloidal tracers. Reversing the electric field allows to reverse the flow and eliminate unwanted drifts. (b) A `vacuum radiometer' with magnetoelectric panes should rotate is the vacuum exerts a pressure on them. According to a prediction by Obukhov \& Hehl (2008) this should not happen. An LED laser on the top of the slab allows to track any displacement of the radiometer in a darkened container (to avoid it being set in motion by ambient light.}\label{expsetup}
\end{figure}

If the Feigel effect has the magnitude we have calculated, it should be possible to test the prediction for a dielectric fluid's velocity using current experimental techniques. Very recently, a $17$ Tesla magnet with temperature control in the range $1.6-300$K and conical ($\pm 10^\circ$) access to its bore has been built (Holmes {\it et al} 2010). The magnet was designed for small angle neutron or X-ray scattering experiments, but it could also used to test our prediction (\ref{predspeed}). Organometallic fluids would be placed in a tube arrangement going in and out of the magnet; a short portion of the tube would reside inside the magnet bore and would be fitted with electrodes to generate the required $10^5\ut{V\,m^{-1}}$ electric fields. A schematic of the set-up is shown in \fig{expsetup}a. As well as electrodes which generate a large electric field, the high strength magnet would have to have a bore large enough to accommodate a tube for the fluid to flow in. In addition, fields need to be uniform over a channel and kept gradient-free to a good approximation. The $17$T magnet described has $0.1$\% uniformity over $1$cm and a bore diameter $\lesssim 4$ cm (A. Holmes, Private Communication), so it would be reasonable to use tubes with $\sim $ mm diameters, as assumed in scenario (a). 
Our prediction is that the vacuum should generate a Poiseuille flow with maximum speed $100\mu$m/s, as obtained from (\ref{PredicPipe}). The corresponding flow rate, $\Phi$ is given by 
\begin{equation}
\Phi=\frac{\pi a^2 U_{max}}{2}\approx 1 \rm{ml\,min}^{-1},
\end{equation}
so that, supposing the fields are as shown in \fig{expsetup}a, experimenters should be able to collect $1$ml of fluid a minute after opening a tap at B (once the fields are in steady state). Reversing one of the fields, the same amount should instead be accumulate at B (this should also allow to subtract out any systematic variations). MMT is transparent, so, if the estimate (\ref{PredicPipe}) is incorrect but a finite vacuum induced flow still exists, it might also be possible to place stripped down (and minimally magnetic!) microscope with a camera down stream to perform particle velocimetry on colloidal tracers. The small flows inside the tube could then be measured downstream far enough away from the magnet to avoid perturbations by large stray fields on the velocimetry apparatus. The mean, $\langle X \rangle$, and mean squared, $\langle X^2 \rangle$, displacements of these tracers could measured as a function of time, $T$:
\begin{eqnarray}\label{md}
&&\langle X \rangle=U_0 T\\
&&\langle X^2 \rangle=4 D T+ U_0^2 T^2\label{msd}
\end{eqnarray}
Integrating measurements over a long enough time and fitting (\ref{md}) and (\ref{msd}) to the data should allow to establish whether there is a nonzero vacuum drift $U_0$ in the fluid and what its magnitude may be. From (\ref{msd}), any vacuum induced drift present can be distinguished beyond doubt from Brownian motion by tracking for $T\gg 4 D/U_0^2$. Supposing tracking can be carried out for for $T_{exp}$, then, if no drift is observed, we infer: $0\le U_0\lesssim\sqrt{4 D/T_{exp}}$. Again, the direction of the flow should be reversed by reversing one of the fields to eliminate other unwanted systematic drifts. 

Turning now to the prediction of Obukhov \& Hehl (2008), this implies that a vacuum radiometer, such as the four pane one shown in \fig{expsetup}b, should not turn. An experimental test would involve a radiometer with thin panes made out of a solid magnetoelectric material (see \fig{expsetup}b). A good candidate for this could be the polar ferrimagnet GaFeO$_3$. Recent studies have characterised the birefringence of this compound in the optical and x-ray ranges (Jung {\it et al} 2004; Kubota {\it et al} 2004). Such studies suggest values larger than those induced by imposed fields in organometallic liquids: $\Delta n\sim 10^{-4}$ (van Tiggelen {\it et al} 2005, van Tiggelen {\it et al} 2006). This should amplify any vacuum effects if they are non-zero. We should point out that $\Delta n\approx \Delta \chi$ drops with increasing temperature for GaFeO$_3$, vanishing with ferrimagnetic order at the transition temperature $T_C\approx 225$ K; further, at low temperatures small magnetic fields are required for a non-zero birefringence (Jung {\it et al} 2004). This means the radiometer should be placed in an evacuated chamber and maintained at temperatures and fields which maximise $\Delta \chi$. Since light could cause the radiometer to turn, the chambers should be light-tight. Light from an LED laser placed on the top of each pane (see \fig{expsetup}b) would then allow a camera in the chamber to track any rotation of the radiometer. A mirror version of the radiometer should also be constructed and its rotation, or lack thereof, tracked. This allows to check the prediction that the rotation should reverse when the direction of the panes' optical axes is reversed, as well as allowing to detect unwanted drifts.

\section{Discussion \label{Discussion}} 

In this paper we argue for the experimental verification of Feigel's theory that the vacuum can transfer momentum to a fluid placed in strong crossed electric. The momentum transfer occurs because vacuum modes in such a fluid are no longer isotropic as in an ordinary dielectric. The crossed electric and magnetic fields $\EE_{\ut{ext}},\BB_{\ut{ext}}$ change the symmetry of the dielectric fluid so that it behaves like a magnetoelectric medium with different refractive indices for waves propagating along or against the direction defined by $\EE_{\ut{ext}}\times\BB_{\ut{ext}}$. Feigel's semi-classical argument allows to derive the net momentum transfer caused by counterpropagating vacuum modes and the corresponding classical contribution due to relative motion in crossed fields. Our alternative derivation which confirms Feigel's result, once minor inaccuracies in the original paper are corrected. Further, a new expression for the vacuum stress on the fluid is derived, predicting a Poiseuille flow in a tube, with maximum speed $U_{max}\approx100\mu$m/s ($2000$ times larger than Feigel's original estimate of $50$nm/s). This prediction contrasts with that of Obukhov \& Hehl (2008) that a magnetoelectric slab in the vacuum experiences no net force. 

Two experiments are proposed to test the above predictions. In Feigel's original scenario of we predict fluid flow which can be measured from the flow rate from an organometallic fluid. Weaker flows could alternatively be measured from particle tracking velocimetry with the same set-up. In the second experiment proposed, Obukhov and Hehl's null prediction for the force on a slab could be tested by measuring the rotation of a vacuum radiometer with panes made of a magnetoelectric material. 

Like all theories, Feigel's theory makes some assumptions which can and should questioned. The boldest assumptions, which we have adopted and are pivotal to our own realistic prediction, are the cut-off frequency assumption and the postulate that vacuum modes see a magnetoelectric (assumptions 4 and 5, respectively, in \sect{Assumptions}). The latter assumption amounts to assuming that the interaction of vacuum modes (virtual photons) with media is identical to that of light (photons). Since vacuum modes are, at least in part, electromagnetic, this seems reasonable; indeed, many QED effects both macroscopic (Casimir-Lifshitz forces) and microscopic (e.g. Lamb and Stark shifts) can be explained by considering the electromagnetic interaction of matter with the vacuum (Milonni 1994). While microscopic approaches can be regularised using a cut-off based on the electron mass, as in the Lamb shift  (Kawka  {\it et al} 2010), Feigel's cut-off assumption has been criticised as improperly regularising the momentum integral, which, it claimed, makes the momentum density incorrectly Lorentz variant. A proper dimensional regularisation, it has been argued, gives a null momentum density, $g_0=0$ (van Tiggelen {\it et al} 2005,  van Tiggelen {\it et al} 2006). However, one could question the fact that the dimensional regularisation employed was pushed beyond the strict limits of its validity, yielding an erroneous result. When evaluating the surface tension of liquid helium films, Schwinger {\it et al} (1978) used a momentum cut-off based on interatomic distances and obtained predictions a factor of 3 larger than experiment. They stated a better agreement would have resulted by employing a microscopic model accounting for the short-distance physics not considered in their continuum model. We believe similar considerations apply to the Feigel effect. As to Lorentz invariance, it is the energy density spectrum, $u_0(\omega)$, which needs to be Lorentz invariant (Boyer 1969). Feigel theory preserves this invariance: since $u_0(\omega)\approx g_0(\omega) c$ we see from (\ref{VacMomentumDens001}) that $u_0(\omega)\sim \omega^3 $.

A quantum microscopic model for the media would allow to properly account for absorption. As mentioned, high frequency modes will not sample a large enough number of atoms to experience a magnetoelectric response (which is the statistical product of many atoms). If this is the case, the vacuum contribution to momentum (pressure) from the absorption of high frequency modes is isotropic and can be neglected. If not, the exact consequences of absorption are hard to fathom in the absence of a microscopic model, but one would intuitively expect a smaller net transfer of momentum (from the inelastic collisions between the virtual photons and the medium, roughly speaking). If the broadband spectrum of magnetoelectric (or effectively magnetoelectric) materials is similar to most dielectric/magnetic materials, then the effect of absorption will be less important for smaller frequencies, which have smaller weight in the momentum density integral. By a similar argument, the dispersion of the real part of the refractive index will matter more at high frequency. A model with a microscopically derived complex, frequency dependent refractive index will dispense of the need for a cut-off (or other regularisation). 

We should also comment on the agreement between Feigel's semi-classical derivation and our own. The latter is also semi-classical in the sense that the form of the expectation value of the momentum density (our starting point) is implicitly given by quantisation of the classical momentum density. The question then arises as to what the appropriate momentum density should be. This is the so-called Abraham-Minkowski problem and, as many researchers have pointed out, it is not a problem at all, provided one is consistent about conserving total momentum (see, e.g., Loudon {\it et al} 2005).  As can be seen from  (\ref{MElagrangian}), Feigel uses the classical pseudo-momentum ${\bf g}=(\DD\times\BB-\EE\times\HH)/4\pi$. We directly quantise this, and evaluate \avg{0|\hat{\gd}|0} using:
\begin{equation}\label{pseudo}
\hat{\gd}=\hat{\gd}_M-\hat{\gd}_A
\end{equation}
where $\hat{\gd}_M=(\hat{\DD}\times\hat{\BB}+\hat{\BB}\times\hat{\DD})/4\pi$ is the Minkowski contribution, where photon momentum is proportional to refractive index $n$, and $\hat{\gd}_A=(\hat{\EE}\times\hat{\HH}+\hat{\HH}\times\hat{\EE})/4\pi$, inversely proportional to it. Feigel on the other hand derives the difference in momentum density for counterpropagating modes classically, and quantises the energy term which emerges. The two derivations thus differ only in the timing of the application of quantisation. Our \avg{0|\hat{\gd}|0} is taken to be proportional to $n_{\kk\lambda}$, which would suggest that we are using the Minkowski term only and not the expectation value of {\ref{pseudo}. Considering the expectation value of the Abraham form vanishes in a vacuum  $\avg{0|\hat{\gd}_A|0} =0$ (van Tiggelen {\it et al} 2005), we see that $\avg{0|\hat{\gd}|0}=\avg{0|\hat{\gd}_M|0}$, i.e. the momentum and its density scales like the refractive index, as we have assumed in our derivations. This Minkoswki pseudo-momentum has been shown to be the appropriate expression for the calculating momentum of matter caused by light (Peierls 1991).

A comparison between the prediction by Obukhov \& Hehl (2008) and that from Feigel's theory is also in order. The two scenarios differ in geometry. In Feigel's suggestion the fluid is unbounded, but in our revision this is true only in the direction along the optical axis. In Obukhov \& Hehl's case, the magnetoelectric slab is bounded by its surfaces and it is the surface contributions which case the net vacuum stress to vanish. In our case, there are no surfaces so the vacuum stress is nonzero. In reality, there will be gradients on each side of the region of the fluid where fields are applied, and it is conceivable that these also might cancel the vacuum stress to vanish. The calculation of these contributions is beyond the scope of this paper, but would represent an interesting matter to pursue theoretically. In their paper Obukhov \& Hehl state that the force on the slab by real photons (e.g. from counter propagating laser beams) should be nonzero. This seems to contradict their vacuum result and they do not make clear why real photons should behave different from virtual ones, though this may well be true. Both the experiments we suggest could be carried out with light and the expectation is that the fluid should move and the radiometer should turn when they are placed in counter propagating laser beams. The difference between real and virtual photons may be that the latter are already involved in providing the magnetoelectric matter with its properties (fine structure etc.) so in some loose sense there are not enough virtual photons to cause a Feigel effect. Another objection might be that if the Feigel effect was real, we would be able to extract small amounts of energy from the vacuum without doing any work (e.g. using the radiometer or Feigel's wheels). It is not clear if second law of thermodynamics applies to the a system macroscopically coupled to the vacuum, since the latter is at zero temperature. On the other hand the third law states that absolute zero cannot be reached in a finite number of steps (Finn 1993), so magnetoelectric media will always have a finite temperature. The vacuum momentum density and stress we have calculated may turn out to be be zero if finite temperature corrections are included. The Casimir effect does not however vanish at finite $T$ (Lamoreaux 2005), so the Feigel effect could be could be simply another weird and wonderful consequence of quantum vacuum fluctuations.

The Feigel effect has inspired many alternative theories, most adopting a regularisable Casimir geometry (van Tiggelen {\it et al} 2006,  Birkeland \& Brevik 2007). Any experimental predictions of these theories are beyond the reach of current instrumentation with Feigel's original prediction recognised to be best candidate for an experimental test (van Tiggelen {\it et al} 2006). Given our realistic prediction is $2000$ times larger than the original, the case for an experimental test is even stronger. Testing Obukhov \& Hehl's null prediction is also compelling. Both predictions involve a qualitative expectation which will hopefully be easy to detect: in one case the fluid moves (or not), in the other the radiometer stays still (or turns). Theories can be challenged theoretically, but the final arbiter of theories in physics is experiment. This author is surprised that, some $7$ years since it was proposed, an experiment to test Feigel's theory has yet to be carried out. We hope this work will stimulate experimentalists to find out if vacuum momentum can be asymmetrically transferred to matter. 

\section{Acknowledgements}

This work was started in 2006 during a visiting scholarship to the Physics Department at Heriot--Watt University. We thank Dr E. Abraham for bringing the Feigel effect to our attention and Dr M. Desmulliez for facilitating the scholarship. We acknowledge discussions with M. de Vries, S. Kotha, P. Frazer and A. Faccia and thank A. Feigel, R. Besseling and A. Holmes for valuable comments. The author gratefully acknowledges support from EPSRC (EP/D073398/1) and the Carnegie Trust.

\appendix{Plane wave modes and refractive indices in a magnetoelectric with orthogonal symmetry\label{AppdxA}}



In this appendix we derive expressions for the amplitudes and refractive
indices of EM modes propagating in a magnetoelectric. The approach is
very similar to the appendix of (Figotin \& Vitebsky, 2001). When describing time-dependent fields in media, the following
Maxwell equations are sufficient (see Chapter 41 of (Schwinger 1998)):

\begin{eqnarray}\label{Max1}
&&\curl{\EE}=-\frac{1}{c}{\pderiv{\BB}{t}}\\
&&\curl{\HH}=\frac{1}{c}{\pderiv{\DD}{t}}\label{Max2}
\end{eqnarray}
In magnetoelectric materials with isotropic permittivity and permeability
tensors ($\hat{\epsilon}=\epsilon \hat{I}$; $\hat{\mu}=\mu \hat{I}$) Equations
(\ref{Max1}), (\ref{Max2}) are supplemented by the constitutive relations (Figotin \& Vitebsky 2001, O'Dell 1970):

\begin{eqnarray}\label{Con1}
&&\DD=\epsilon \EE + \X \HH\\
&&\BB=\mu \HH + \Xt \EE \label{Con2}
\end{eqnarray}
where the $\X$ is the magnetoelectric susceptibility tensor, defined, in matrix form, by:
\begin{equation}\label{MESuscTensor0}
\X\equiv \left(
                           \begin{array}{ccc}
                             0 & \chixy & 0 \\
                             \chiyx & 0 & 0 \\
                             0 & 0 & 0 \\
                           \end{array}
                           \right)
\end{equation}
In (\ref{MESuscTensor}), $\chixy, \chiyx$ are magnetoelectric susceptibilities. They are a measure of
the electrical polarisation caused by a magnetic field and the magnetisation caused by an electric
fields applied along the $x$ and $y$ axes in the magnetoelectric.
%
%
We assume plane wave solutions propagating in the z-direction:

\begin{equation}\label{WaveComp}
\EE_{\kk\lambda}(z,t)=E_{0\kk}\exp{i(k_\lambda z -\omega t)}\ee_{\kk\lambda}\equiv
E_{\kk\lambda}\ee_{\kk\lambda}
\end{equation}
with wave-vector $\kk_\lambda$, frequency $\omega$ and polarisation vector
$\ee_{\kk\lambda}$. For such waves (removing the suffices for clarity):

\begin{equation}
\curl{\EE} =\partial_z\left(
                          \begin{array}{c}
                            -E_y \\
                            E_x \\
                            0 \\
                          \end{array}
                        \right)
                        =\partial_z\left(
                           \begin{array}{ccc}
                             0 & -1 & 0 \\
                             1 & 0 & 0 \\
                             0 & 0 & 0 \\
                           \end{array}
                         \right)
                         \left(
                          \begin{array}{c}
                            E_x \\
                            E_y \\
                            0 \\
                          \end{array}
                        \right)
\end{equation}
Thus, defining:

\begin{equation}
\sig\equiv \left(
                           \begin{array}{ccc}
                             0 & -1 & 0 \\
                             1 & 0 & 0 \\
                             0 & 0 & 0 \\
                           \end{array}
                           \right)
\end{equation}
We can write:

\begin{equation}
\curl{\EE} =\sig \partial_z \EE
\end{equation}
Analogous considerations apply to $\curl{\HH}$. Thus, since $\partial_z{\EE}=i
k \vect{E}$ and $\partial_t{\BB}=- i \omega \BB$ and similarly for $\HH$ and
$\DD$, Equations (\ref{Max1}) become:

\begin{eqnarray} \label{Max21}
n \sig \EE&=& \BB\\
n \sig \HH&=&-\DD \label{Max22}
\end{eqnarray}
where we have defined $n\equiv k (c/\omega)$. To find the values of $k$ for
which the plane waves are solutions of Maxwell's equations.

Substituting (\ref{Con1}) and (\ref{Con2}) into (\ref{Max21}) and (\ref{Max22}) and
rearranging we find:

\begin{eqnarray} \label{Max31}
&&\HH= \frac{1}{\mu}(n\sig-\Xt)\EE\\
&&(n\sig+\X)\HH= -\epsilon \EE \label{Max32}
\end{eqnarray}
Substituting for $\HH$ from (\ref{Max31}) into (\ref{Max32}) we find the following
eigenvalue equation:
\begin{equation}\label{eigen}
\hat{N}\EE = \epsilon \mu\EE
\end{equation}
where:
\begin{equation}\label{MESuscTensor}
\hat{N}\equiv-(n\sig+\X)(n\sig-\Xt)=\left(
                                      \begin{array}{ccc}
                                        (n-\chixy)^2 & 0 & 0 \\
                                        0 & (n+\chixy)^2 & 0 \\
                                        0 & 0 & 0 \\
                                      \end{array}
                                    \right)
\end{equation}
\subsection{Mode Amplitudes and Indices of Refraction}

We look for linearly polarised plane wave solutions to Maxwell's equations.
There are two independent polarisations. Consider first plane wave with $\EE$
along $\ee_1=(1,  0,  0)$. The eigenvalue equation (\ref{eigen}) then entails:
\begin{equation}
                                    \left(
                                      \begin{array}{ccc}
                                        (n_1-\chixy)^2 & 0 & 0 \\
                                        0 & (n_1+\chixy)^2 & 0 \\
                                        0 & 0 & 0 \\
                                      \end{array}
                                    \right)
\left(
  \begin{array}{c}
    1 \\
    0 \\
    0 \\
  \end{array}
\right)
                                    =
                                    \epsilon \mu
                                    \left(
  \begin{array}{c}
    1 \\
    0 \\
    0 \\
  \end{array}
\right)
\end{equation}
So that the refractive indices for waves propagating in the $\kk$ and $-\kk$
directions with polarisation 1 are given by:
\begin{eqnarray}
n_{\kk,1}&=&\sqrt{\epsilon \mu}+\chixy\\
n_{-\kk,1}&=&-\sqrt{\epsilon \mu}+\chixy
\end{eqnarray}
Analogously, for waves with $\EE$ along $\ee_2=(0, 1, 0)$ the eigenvalue
equation (\ref{eigen}) requires:
\begin{eqnarray}
n_{\kk,2}&=&\sqrt{\epsilon \mu}-\chiyx\\
n_{-\kk,2}&=&-\sqrt{\epsilon \mu}-\chiyx
\end{eqnarray}
The magnetic field components corresponding to the above electric field modes
are found using the Maxwell equation (\ref{Max21}). For polarisation $\ee_1$:
\begin{equation}
\BB_1=n_1 \sig \EE_1=\left(
                           \begin{array}{ccc}
                             0 & -n_1 & 0 \\
                             n_1 & 0 & 0\\
                             0 &   0 & 0\\
                           \end{array}
                         \right)
                         \left(
                          \begin{array}{c}
                            1 \\
                            0 \\
                            0\\
                          \end{array}
                        \right)E_1
                        =
                        \left(
                          \begin{array}{c}
                            0 \\
                            n_1 \\
                            0 \\
                          \end{array}
                        \right)E_1
\end{equation}
and similarly for $\ee_2$:
\begin{equation}
\BB_2=n_2 \sig \EE_2=\left(
                          \begin{array}{c}
                            -n_2 \\
                            0 \\
                            0\\
                          \end{array}
                        \right)E_2
\end{equation}
So finally, the four possible modes of propagation in a magnetoelectric are:
\begin{eqnarray}\label{Modes}
&&[\EE_{\kk1}, \BB_{\kk1}]=E_{\kk1}[\ee_{1}, n_{\kk1}\ee_{2}],\,\,[\EE_{-\kk1}, \BB_{-\kk1}]=E_{-\kk1}[\ee_{1}, n_{-\kk1}\ee_{2}], \\
&&[\EE_{\kk2}, \BB_{\kk2}]=E_{\kk2}[\ee_{2}, -n_{\kk2}\ee_{1}],\,\,[\EE_{-\kk2}, \BB_{-\kk2}]=E_{-\kk2}[\ee_{2}, -n_{-\kk2}\ee_{1}]
\end{eqnarray}
or, in Feigel's notation $(1/E_{\kk\lambda})(E_x, E_y, B_x, B_y)$:
\begin{eqnarray*}\label{RefractiveIndices}
&&(1, 0, 0, n_{\kk 1}),\,\,\,\, (0, 1, -n_{\kk 2}, 0),\\
&&(1, 0, 0, n_{-\kk 1}), \,\,\,\,(0, 1, -n_{-\kk 2}, 0).
\end{eqnarray*}

\appendix{Lagrangian and Hamiltonian densities for magnetoelectric media \label{AppdxB}}

In general the electromagnetic Lagrangian in ponderable media is given by:
\begin{equation}
L=\iint \Lag d\vect{r}\,dt
\end{equation}
where the Lagrangian density is given by:
\begin{equation}\label{EMLagMedia}
\Lag=\frac{1}{4 \pi}\left[\frac{1}{2}\EE\cdot\DD-
\frac{1}{2}\BB\cdot\HH\right]
\end{equation}
A magnetoelectric material satisfies the constitutive relations (\ref{Con1}) and (\ref{Con2}).
These give $\DD=\DD(\EE,\HH)$ and $\BB=\BB(\EE,\HH)$. We can invert (\ref{Con1})
to find $\HH=\HH(\EE,\BB)$ and so $\DD=\DD(\EE,\BB)$ from (\ref{Con2}):
\begin{eqnarray}\label{Con11}
&&\HH=\frac{1}{\mu}\left( \BB - \Xt \EE \right)\\
&&\DD=\epsilon \EE + \frac{1}{\mu}\X \left( \BB - \Xt \EE \right) \label{Con22}
\end{eqnarray}
Substituting (\ref{Con11}) and (\ref{Con22}) into the Lagrangian density \ref{EMLagMedia}, we find:
\begin{equation}\label{EMLagMedia2}
\Lag=\frac{1}{4 \pi}\left[\frac{1}{2}\EE\cdot\left(\epsilon \EE + \frac{1}{\mu}\X  \BB -
\frac{1}{\mu}\X\Xt \EE \right)-
\frac{1}{2}\BB\cdot\left( \frac{1}{\mu}\BB - \frac{1}{\mu}\Xt \EE \right)\right]
\end{equation}
or
\begin{equation}\label{EMLagMedia3}
\Lag=\frac{1}{4 \pi}\left[\frac{1}{2}\epsilon\EE^2-\frac{1}{2\mu}\BB^2+\frac{1}{2 \mu}\EE\cdot (\X  \BB) +
 \frac{1}{2 \mu}\BB \cdot(\Xt \EE) -\frac{1}{2 \mu} \EE \cdot(\X\Xt  \EE) \right]
\end{equation}
Letting $Q=\EE\cdot (\X  \BB)$ we see $Q$ is a scalar ($Q=Q^T$) so that $\EE\cdot (\X  \BB)=\BB \cdot(\Xt \EE)$ and, neglecting terms of order $\chixy^2,\chiyx^2$:
\begin{equation}\label{EMLagMedia4}
\Lag=\frac{1}{4 \pi}\left[\frac{1}{2}\epsilon\EE^2-\frac{1}{2\mu}\BB^2 +
 \frac{1}{\mu}\BB \cdot(\Xt \EE) \right] + o(\chi^2)
\end{equation}
Similarly, the Hamiltonian is given by:
\begin{equation}\label{EMHam}
H=\frac{1}{4 \pi}\left[\frac{1}{2}\EE\cdot\DD+
\frac{1}{2}\BB\cdot\HH\right]\approx \frac{1}{4 \pi}\left[\frac{1}{2}\epsilon\EE^2+\frac{1}{2\mu}\BB^2 \right]+ o(\chi^2)
\end{equation} 

\appendix{Inaccuracies in Feigel's original derivation\label{AppdxC}}

\begin{table}
\begin{center}
\begin{tabular}{l l l}
\hline
{\bf quantity} & \vline\, {\bf Feigel's value} &  \vline\, {\bf correct value}\\
\hline
   & \vline\, $(1, 0, 0, \sqrt{\epsilon \mu})$& \vline\, $(1, 0, 0, \sqrt{\epsilon \mu}+\chixy)$\\
magnetoelectric modes & \vline\, $(1, 0, 0, -\sqrt{\epsilon \mu})$& \vline\, $(1, 0, 0, -\sqrt{\epsilon \mu}+\chixy)$\\  
& \vline\, $(0, 1, -\sqrt{\epsilon \mu}, 0)$& \vline\, $(0, 1, -\sqrt{\epsilon \mu}+\chiyx, 0)$\\  
& \vline\, $(0, 1, \sqrt{\epsilon \mu}, 0)$& \vline\, $(0, 1, \sqrt{\epsilon \mu}+\chiyx, 0)$\\
\hline
momentum density of mode $\kk$ & \vline\,$2 \Delta \chi \frac{1}{c} (\epsilon+\frac{1}{\mu}) \frac{E^2_{0 k}}{4 \pi}  $ &\vline\,  $2\Delta\chi \frac{1}{c}\frac{\epsilon E^2_{0 k}}{4\pi}$\\
\hline
vacuum momentum density expectation  & \vline\, $ \frac{1}{32 \pi^3} \Delta \chi  \frac{1+\epsilon \mu}{\mu}\frac{\hbar \omega_{max}^4}{c^4}$ & \vline\, $ \frac{1}{8 \pi^2} \Delta \chi  \frac{\hbar \omega_{max}^4}{c^4}$\\
\hline
\end{tabular}
\end{center}
\caption{Trivially incorrect quantities in Feigel's derivation (Feigel 2004) and the correct values derived in this paper.} \label{Inaccuracies}
\end{table}
%

%
%









\end{document}